%% file: 2018-RTSS-MAC.tex
\documentclass[10pt,conference,compsocconf]{IEEEtran}  
\IEEEoverridecommandlockouts

\usepackage{xspace}
\usepackage{amsmath, amssymb}
\usepackage{mathtools}
\usepackage{mathptmx}
\usepackage{nicefrac}
\usepackage{booktabs}
\usepackage{psfrag}
\usepackage{listings}
\usepackage{tabularx}
\usepackage[para,online,flushleft]{threeparttable}
\usepackage{multirow}
\usepackage{tikz,pgfplots}
\usetikzlibrary{pgfplots.groupplots}
\usepackage{url}

\usepackage{enumitem}

\usepackage[nolist,nohyperlinks]{acronym}
\acrodef{WSN}[WSN]{Wirless Sensor Network}

\usepackage{tikz,xstring,siunitx,pgfplots,psfrag}
\usepackage{pgfplotstable}
\usepackage{ifthen}

\usetikzlibrary{shapes,arrows}
\usetikzlibrary{calc,patterns,decorations.pathmorphing,decorations.markings}
\usetikzlibrary{positioning,automata}
\usetikzlibrary{pgfplots.groupplots}

\tikzstyle{block} = [draw, rectangle, minimum height=1em, minimum width=1em]
\tikzstyle{sum} = [draw, circle, node distance=1.5cm]
\tikzstyle{input} = [coordinate]
\tikzstyle{output} = [coordinate]

\usepackage{color}

\definecolor{mygreen}{rgb}{0,0.6,0}
\definecolor{mygray}{rgb}{0.5,0.5,0.5}
\definecolor{mymauve}{rgb}{0.58,0,0.82}

\usepackage{listings}
\usepackage[colorinlistoftodos, textwidth=20mm, textsize=footnotesize]{todonotes}

\begin{document}
\onecolumn

\section*{Preprint version}
F. Terraneo, P. Polidori, A. Leva, W. Fornaciari "TDMH-MAC: Real-time and multi-hop in the same wireless MAC" IEEE Real-Time Systems Symposium (RTSS), Nashville, USA, December 2018\\
\twocolumn
\clearpage

\title{\LARGE \bf TDMH-MAC: Real-time and multi-hop in the same wireless MAC
}

\author{\IEEEauthorblockN{Federico Terraneo\IEEEauthorrefmark{1},
                          Paolo Polidori\IEEEauthorrefmark{2},
                          Alberto Leva\IEEEauthorrefmark{1},
                          William Fornaciari\IEEEauthorrefmark{1},
                         }
        \IEEEauthorblockA{\IEEEauthorrefmark{1}Politecnico di Milano, Italy.
                           Email: \{federico.terraneo,alberto.leva,william.fornaciari\}@polimi.it             
                         }
        \IEEEauthorblockA{\IEEEauthorrefmark{2}Former graduate student at the Politecnico di Milano. Email: paolo.polidori@mail.polimi.it
                         }
}

\maketitle
\thispagestyle{empty}
\pagestyle{empty}

\begin{abstract}
\input{./sections/00-Abstract.tex}
\end{abstract}

\acresetall

\input{./sections/01-Introduction.tex}
\input{./sections/02-RelatedWork.tex}

\input{./sections/03-ProposedProtocol.tex}

\input{./sections/04-TopologyCollection.tex}

\input{./sections/05-ScheduleDistribution.tex}

\input{./sections/06-DataTransmission.tex}

\input{./sections/07-SimScalability.tex}

\input{./sections/08-ExpRes.tex}

\input{./sections/09-Conclusions.tex}


\end{document}

%% file: sections/00-Abstract.tex
Supporting real-time communications over Wireless networks (WSNs) is a tough challenge, due to packet collisions and the non-determinism of common channel access schemes like CSMA/CA. Real-time WSN communication is even more problematic in the general case of multi-hop mesh networks. For this reason, many real-time WSN solutions are limited to simple topologies, such as star networks. We propose a real-time multi-hop WSN MAC protocol built atop the IEEE 802.15.4 physical layer. By relying on precise clock synchronization and constructive interference-based flooding, the proposed MAC builds a centralized TDMA schedule, supporting multi-hop mesh networks. The real-time multi-hop communication model is connection-oriented, using guaranteed time slots, ad enables point-to-point communications also with redundant paths. The protocol has been implemented in simulation using OMNeT++, and the performance has been verified in a real-world deployment using Wandstem WSN nodes.

%% file: sections/01-Introduction.tex
\section{Introduction}
\label{sec:Intro}

Wireless protocols supporting real-time communications are an open research area with wide applications in the industry. Current trends such as Industry 4.0 pose wireless communication at the forefront in drastically increasing the automation level of production factories~\cite{7945906,7993826}.
However, plant-wide industrial control introduces both real-time requirements on communications as well as the need for multi-hop protocols to cope with the large scale, dense deployment of sensing and actuation devices. Moreover, wireless sensors are often battery operated in order not to require a wired power infrastructure, which calls for energy efficient wireless protocols.

To address these issues we introduce a new wireless MAC (Medium Access Control) protocol, based upon the 802.15.4 physical layer, capable of supporting real-time multi-hop commuinications. We name our protocol TDMH-MAC, where TDMH stands tor Time Deterministic Multi-Hop.

TDMH-MAC builds upon the recent availability of sub-$\mu$s clock synchronization schemes with negligible energy overhead~\cite{bib:TerraneoEtAl-2014a}, which allowed to reverse the mainstream approach of basing communication timing and synchronization on an already running MAC protocol. Instead, we assume clock synchronization as the foundation of our MAC, and exploit it fully for all subsequent operations, such as data packet scheduling, as well as collection of the network topology and route dissemination. Said otherwise, alternative approaches \emph{first} bring up the network, and \emph{then} seek clock synchronisation and care about timing. On the contrary, with our approach, the nodes synchronize their clocks \emph{before the network is formed}. The synchronization is then exploited to build a centralized TDMA (Time Division Multiple Access) MAC which eliminates the time uncertainty caused by statistical multiplexing schemes such as CSMA/CA (Carrier Sense Multiple Acces with Collision Avoidance).

In a centralized mesh MAC, constructing the graph of the network as well as transmitting it to the coordinator represents a problem, as much as efficiently disseminating the route information computed by the coordinator to all the nodes in the network. We solve the first problem through an efficient topology collection scheme, which is one of the key innovations of TDMH-MAC. The problem of route dissemination is instead handled using constructive interference flooding to minimize network management overhead.

The result is a MAC with the following innovative characteristics:
\begin{itemize}
\item it is based on a mesh topology, allowing redundant paths for reliability and load balancing,
\item the network topology is constantly updated allowing for adaptation and robustness to link and node failure,
\item thanks to a centralized scheduling of packet routes, it provides collision-free deterministic communication,
\item a connection-oriented model provides guaranteed allocation of resources to data streams,
\item data delivery follows a periodic model with guaranteed period and bounded latency, making the MAC suitable for networked control systems,
\item communication streams can be opened at runtime from any node to any node.
\end{itemize}

%% file: sections/02-RelatedWork.tex
\section{Related Work}
\label{sec:RelatedWork}

IEEE 802.15.4 is the current standard for low power radio communication.
Many MAC protocols have been built on this standard by the Wireless Sensor Network (WSN) community, a taxonomy of which can be found in~\cite{wsnmacsurvey}.

Asynchronous protocols use duty cycling to reduce receivers' energy consumption, but do not synchronize the nodes' sleep cycles, resulting in aloha-like protocols.
B-MAC~\cite{bmac} is an example. These protocols suffer from high latency and poor time determinism caused by the use of a CSMA/CA scheme with long channel sensing intervals.

Synchronous protocols group nodes in clusters with a common sleep/wake schedule, enhancing throughput and reducing delay at the cluster level. An example is S-MAC~\cite{smac}.
Routing packets between clusters is however still affected by high latencies, a problem partially addressed by enhancements such as AS-MAC~\cite{asmac} and DMAC~\cite{dmac}.

Frame-slotted protocols further enhance performance by a fine enough synchronization to make TDMA possible. TreeMAC~\cite{treemac} uses a gathering tree structure to collect data centrally, defines \emph{frames} composed of 3 slots, and distributes frames among nodes at the same depth, dividing the frame space assigned to their parent.
An improvement is PackMAC~\cite{packmac}, that in parallel to TreeMAC runs a distributed free time-slot search algorithm.
Other works focus on maximizing channel utilization under high contention by reusing free TDMA slots for CSMA, such as in TRAMA~\cite{trama}.

Multichannel protocols reduce contention and improve throughput. MMSN~\cite{mmsn} uses a default channel to coordinate data transmissions, which are then performed on a dynamically assigned one. GBCA~\cite{gbca} uses a game theoretic approach to minimize interference, a problem shown to be NP-hard. In other protocols channels are assignmed by the sender~\cite{mclmac} or the receiver~\cite{ymac}. MuChMAC~\cite{muchmac} performs channel hopping using a pre-shared sequence. Our MAC can be extended to the multichannel case, but this is outside the scope of this paper.

Time Slotted Channel Hopping (TSCH) combines TDMA and multichannel support. It is based on a cluster-tree topology where time is divided in repeating slotframes, composed of time slots where data transmission occurs. Network management and synchronization is performed through the transmission of Enhanced Beacons (EBs). TSCH requires transmissions to be scheduled for proper operation, however no scheduler is present in the specification, as this is thought as a point of customization. Two different families of scheduling algorithms have been proposed. Centralized schedulers include Traffic Aware Scheduling Algorithm (TASA)~\cite{tasa}, which starts with a statically configured topology represented in a graph structure, where scheduling is extracted via matching and vertex coloring. Distributed ones include DeTAS~\cite{dtas}, a randomized load balancer~\cite{sdswcd}, and DIVA \cite{diva}, that tries not to concentrate traffic towards the root node, preferring peripheral paths.

Deterministic and Synchronous Multi-Channel Extension (DSME)~\cite{OpenDSME} exploits the Collision Free Part (CFP) to reduce collisions.
Beacon scheduling and slot allocation are managed in a distributed way. This means that each node can autonomously allocate or deallocate slots, resulting in a different architecture with respect to TSCH. Low Latency Deterministic Network (LLDN) is specifically designed for low latency real-time applications. Its goal is sampling and data collection every 10ms from 20 different sensors. However, to achieve this performance on a low data rate network like 802.15.4, it only supports a star topology~\cite{7005204}.

There is also a number of commercial protocols developed upon 802.15.4, whose specifications are not openly available.
ZigBee is perhaps the most common, but is not targeted to real-time applications. Linear Technologies has developed a commercial version of TSCH, called SmartMesh IP®\cite{smartmeship}. WirelessHART\cite{wirelesshart} and ISA-100.11a~\cite{ISA-100.11a} are two closed industry standards. As illustrated in \cite{whartvsisa} they both provide a TDMA protocol with frequency hopping, use a mesh topology and have limits of thousands of devices, though with large networks consumptions and latencies may grow unpredictably. Unlike TDMH-MAC, none of these protocols rely on constructive interference flooding.

Real-time wireless networks can also adopt other standards than 802.15.4.
Bluetooth Low Energy has recently been proposed~\cite{8355905} for real-time multi-hop networks, achieving promising results for a protocol stack not originally intended for this purpose.
RT-WiFi~\cite{6728869} transmits TDMA-scheduled packets enjoying the higher data rate of Wi-Fi, but those networks are limited to a star topology. Other approaches to real-time Wi-Fi networks try to cope with the limitations of CSMA, by either modifying the backoff algorithm and using Quality of Service (QoS)~\cite{7389381} to prioritize real-time traffic, or to jam the channel to first to stop non real-time traffic and then transmitting, a technique refered to as bandjacking~\cite{8336984}.

Table~\ref{tab:mac-comparison} compares the features of our MAC to the most relevant protocols whose specifications are openly available.

\begin{threeparttable}[b]
\vspace{4mm}
\caption{Comparison of existing wireless MAC protocols.}
{\footnotesize \centering
		\begin{tabularx}{\columnwidth}{l|ccccc}
			Feature             & TDMH-MAC      & TSCH         & DSME       & LLDN       & rt-WiFi\\
			\hline
			Multi-hop           & \checkmark & \checkmark   & \checkmark &            &  \rule{0pt}{2.6ex}\\ [1mm]
				Guaranteed          & \multirow{2}{*}{\checkmark} & \multirow{2}{*}{} & \multirow{2}{*}{} & \multirow{2}{*}{\checkmark} & \multirow{2}{*}{\checkmark} \\
                period \\ [1mm]
				Spatial             & \multirow{2}{*}{\checkmark} & \multirow{2}{*}{} & \multirow{2}{*}{} & \multirow{2}{*}{} & \\
				redundancy \\ [1mm]
				Temporal            & \multirow{2}{*}{\checkmark} & \multirow{2}{*}{\checkmark} & \multirow{2}{*}{feasible} & \multirow{2}{*}{feasible} & \multirow{2}{*}{\checkmark} \\
				redundancy \\ [1mm]
			Management          & C\tnote{1} & C\tnote{1}/D\tnote{2} & D\tnote{2} & C\tnote{1} & C\tnote{1} \\ [1mm]
			Topology            & mesh       & ct\tnote{3} & ct\tnote{3} & star & star \\ [1mm]
		\end{tabularx}
		\vspace{0.4em}
		\begin{tablenotes}
            \item[1] centralized \item[2] distributed \item[3] cluster-tree
        \end{tablenotes}
        \vspace{0.4em}
	\label{tab:mac-comparison}
}
\end{threeparttable}

As can be seen from Table~\ref{tab:mac-comparison}, existing protocols can be divided between those guaranteeing tight latency bounds, which are however limited to star topologies, and protocols supporting multi-hop networks, where providing any form of latency bound is a much more uncommon feature.
For what concerns the latter, network topology is usually limited to a cluster-tree to overcome the difficulties in discovering the topology and keeping it updated.
Such a solution however removes a-priori some links which could be used for spatial redundancy or load balancing. TDMH-MAC innovates in this respect by providing an efficient solution to the topology collection problem.
Moreover, when considering centralized protocols, TDMH-MAC is, to the best of the authors' knowledge, the first using constructive interference flooding to disseminate routing information, a solution which is known to be very efficient in terms of channel usage~\cite{bib:FerrariEtAl-2011a}, thus reducing control overhead and leaving more radio time for data exchange.
Although it is difficult to compare TDMH-MAC with other protocols due to their very different nature, the flexibility and efficiency of TDMH-MAC can be expected to provide tighter latency bounds, also in complex multi-hop scenarios, despite the data rate limitations of 802.15.4.

%% file: sections/03-ProposedProtocol.tex
\section{TDMH-MAC Protocol Design}
\label{sec:Protocol}

TDMH-MAC is a centralized, connection-oriented TDMA mesh protocol. Data communication is performed through streams, which are logical point-to-point links between nodes. Streams can be dynamically opened between any two nodes in the network. By the design of the protocol, streams can not only have dedicated bandwidth, but also a dedicated \emph{period} between application-level packets, allowing to support real-time applications such as control loops. Application-level packets are routed with a bounded end-to-end latency.
To enhance reliability, individual application-level packets can be transmitted through multiple frames over the wireless links for redundancy. Moreover, the mesh network topology permits to transmit frames also through different paths for increased reliability in case of link or node failure.

Being a centralized protocol, one node in the network is assigned the master role. The master node periodically sends clock synchronization frames, allowing other nodes to synchronize and join the network. As in TDMH-MAC nodes cannot join the network before they are synchronized, we use the FLOPSYNC-2~\cite{bib:TerraneoEtAl-2014a} clock synchronization scheme as it can operate also in the absence of a running  MAC, thus solving the initial synchronization problem before the MAC is operational. This clock synchronization scheme compensates for nonlinear clock skew caused by temperature variations without requiring temperature measurements, and synchronizes a network to below one microsecond despite transmitting only one synchronization frame every 10 to 60 seconds, thus achieving a very low energy and bandwidth overhead.

All the nodes in the network periodically send their knowledge of the network topology to the master node, allowing it to have a full graph of the network. This task is called topology collection.
From this information, the master node can globally schedule the real-time communication streams, with the result of a collision-free data transmission. The computed schedule is then flooded through the network.
The topology collection is performed continuously, so that the master is kept up to date with the current network graph and can trigger a reschedule should a node fail, or a link become unreliable due to environmental changes or external interference.
The schedule is instead updated and flooded only when needed.
It is expected that the master role is assigned to the network gateway, which connects the network to the Internet or, in an industrial control plant, to a Programmable Logic Controller (PLC) or Supervisory Control And Data Acquisition (SCADA) system. In these kind of networks the gateway is already a single point of failure, so the centralized protocol does not add another failure mode.
For use cases where the network can continue to function also if isolated, an election procedure can be added to preserve operation should the master fail, but this is outside the scope of this work.

The proposed MAC is not targeted at use cases with mobile nodes, as it assumes that topology changes are infrequent enough to allow them to be detected by the topology collection and a new schedule to be distributed with negligible downtime.

The use of multiple paths for redundancy also helps in preventing downtimes due to small topology changes, as a link or node failure in one of the paths would be masked by the other paths until a new schedule takes effect.

\subsection{TDMH-MAC activities}

The TDMH-MAC protocol can be logically viewed as composed of three distinct activities, control downlink, control uplink and data transmission.
The control downlink activity is performed using the Glossy~\cite{bib:FerrariEtAl-2011a} constructive interference flooding scheme. Flooded frames all originate from the master node. The purpose of this activity is to distribute updates to the schedule used for data transmission, as well as clock synchronization frames. The use of a dedicated flooding scheme to efficiently disseminate network management information to all the nodes in the network is a distinctive characteristic of TDMH-MAC.

The control uplink activity is used for on-line topology collection, and for opening new communication streams between nodes. It is performed through a round robin scheme where each node in turn both informs other nodes of its existence, and transmits its knowledge of the network topology towards the master node. The proposed topology collection scheme is one of the key innovations of the protocol, and relies on information from both the clock synchronization and flooding scheme to efficiently gather information about the network topology and propagate it to the master node.

The data transmission activity is where application data are transmitted according to the global TDMA schedule. Time is divided in slots, where data frames are transmitted.
The operation of the MAC in the data slots is entirely schedule-driven, where a node knows in advance when to transmit, receive or sleep to save energy.

\subsection{Structure of TDMH-MAC}

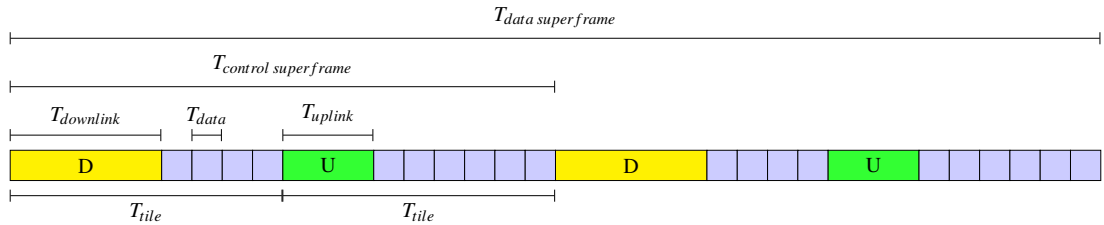
\begin{figure*}[tb] 
 \begin{center}
  \resizebox{1.65\columnwidth}{!}{\input{./img/MACstructure.tex}}
 \end{center}
 \caption{Temporal organization of the TDMH-MAC protocol. Transmission is organized in tiles of two types but of equal length. Downlink tiles start with a flooded frame from the master, Uplink tiles start with a control uplink slot for topology collection and stream opening requests.}
 \label{fig:ProtocolStructure}
\end{figure*}

TDMH-MAC is temporally organized as shown in Figure~\ref{fig:ProtocolStructure}. The protocol is organized in \emph{tiles}. All tiles begin with a control slot, where one control transmission, either downlink or uplink, occurs. The rest of the tile is occupied by data slots where data frames
are transmitted. All tiles are the same length, and since downlink slots are larger due to the need to flood a frame across multiple hops, downlink tiles have fewer data slots.

The shortest repeating sequence of downlink and uplink tiles is called a control superframe. A control superframe must have at least one downlink and one uplink tile for the MAC to be able to perform both activities. In a common configuration the control superframe is composed of exactly one downlink and one uplink tile, but other configurations are possible to modify the downlink/uplink ratio in order to optimize the MAC for specific use cases. For example, in a network where streams are frequently opened and closed by the master node, it is possible to increase the ratio of downlink to uplink slots to disseminate schedules faster.

The schedule computed by the master node defines the data superframe, whose length is necessarily a multiple of the control superframe due to the asymmetry in the number of data slots of the different tiles. 
In each data slot, each node performs the operation prescribed by the TDMA schedule, which can be one of the following five operations: transmit a frame containing data from upper layers, receive a frame and store it in a buffer to later forward it, forward a buffered frame, receive a frame and pass it to upper layers, or sleep.

\subsection{Network configuration}

The network configuration is a set of parameters that all nodes need in order to join a network. Except for the node ID that is unique for each node, all the other parameters are the same for all nodes in a network. The network configuration is defined when the network is created and first stored in the master node. Each node, to be able to connect to the network, needs to perform an association procedure with the master node. During association the node ID is assigned and the network configuration is shared through an out-of-band communication channel.

The network configuration can be divided in four parts: general configuration, tile configuration, topology configuration and node ID.
The general configuration contains the network PAN ID and radio channel as specified by IEEE 802.15.4, as well as the FLOPSYNC-2 clock synchronization period and information on whether propagation delay compensation should be preformed~\cite{bib:TerraneoEtAl-2015a}.
The tile configuration contains the tile duration in milliseconds, the control superframe structure and data transmission configuration (data slot length, data frame size).
The topology configuration contains the maximum number of hops the network can have, which determines the downlink slot length, and the maximum number of nodes the network can have.
The node ID is an integer number starting from zero for the master node, ranging up to the maximum number of nodes minus one.

%% file: img/MACstructure.tex
\begin{tikzpicture}

\node[right,draw=black,fill=yellow,
      minimum width=25mm,minimum height=5mm]
      at (0,0) (D1) {D}; 
\foreach \x in {25mm,30mm,35mm,40mm}
{
\node[right,draw=black,fill=blue!20!white,
      minimum width=5mm,minimum height=5mm]
      at (\x,0){};
}
\node[right,draw=black,fill=green!80!white,
      minimum width=15mm,minimum height=5mm]
      at (45mm,0) (U1) {U}; 
\foreach \x in {15mm,20mm,25mm,30mm,35mm,40mm}
{
\node[right,draw=black,fill=blue!20!white,
      minimum width=5mm,minimum height=5mm]
      at (\x+45mm,0){};
}

\node[right,draw=black,fill=yellow,
      minimum width=25mm,minimum height=5mm]
      at (90mm,0) (D2) {D}; 
\foreach \x in {25mm,30mm,35mm,40mm}
{
\node[right,draw=black,fill=blue!20!white,
      minimum width=5mm,minimum height=5mm]
      at (\x+90mm,0){};
}
\node[right,draw=black,fill=green!80!white,
      minimum width=15mm,minimum height=5mm]
      at (135mm,0) (U2) {U}; 
\foreach \x in {15mm,20mm,25mm,30mm,35mm,40mm}
{
\node[right,draw=black,fill=blue!20!white,
      minimum width=5mm,minimum height=5mm]
      at (\x+135mm,0){};
}

\draw[|-|] (0,-5mm)
   -- node[pos=0.5,below]{$T_{tile}$}
  (45mm,-5mm);
\draw[|-|] (45mm,-5mm)
   -- node[pos=0.5,below]{$T_{tile}$}
  (90mm,-5mm);
\draw[|-|] (0,5mm)
   -- node[pos=0.5,above,yshift=0.5mm]{$T_{downlink}$}
  (25mm,5mm);
\draw[|-|] (30mm,5mm)
   -- node[pos=0.5,above,yshift=0.5mm]{$T_{data}$}
  (35mm,5mm);
\draw[|-|] (45mm,5mm)
   -- node[pos=0.5,above,yshift=0.5mm]{$T_{uplink}$}
  (60mm,5mm);
\draw[|-|] (0,13mm)
   -- node[pos=0.5,above,yshift=0.5mm]{$T_{control\;superframe}$}
  (90mm,13mm);
\draw[|-|] (0,21mm)
   -- node[pos=0.5,above,yshift=0.5mm]{$T_{data\;superframe}$}
  (180mm,21mm);

\end{tikzpicture}

%% file: sections/04-TopologyCollection.tex
\section{Network Connection and Topology Collection}
\label{sec:TopologyCollection}

In order for the master node to be able to schedule streams, knowledge about the network graph is required.
In TDMH-MAC, this is achieved through a topology collection distributed algorithm which builds upon the FLOPSYNC-2 clock synchronization scheme and Glossy flooding scheme.

FLOPSYNC-2~\cite{bib:TerraneoEtAl-2014a} is a clock synchronization scheme that uses control theory to compensate for the nonlinear clock skew of each node resulting in sub-$\mu$s clock synchronization with a low energy and bandwidth overhead. Although the original FLOPSYNC-2 did not put timestamps in synchronization frames, TDMH-MAC extends those frames by including a 32 bit counter starting at 0 and incremented at every synchronization period. Since every node knows the synchronization period, by multiplying it by the counter it is possible to know the global network time, which is used to schedule every network activity, including the control uplink which is where the topology collection is performed.

FLOPSYNC-2 frames are transmitted through constructive interference flooding with the use of the Glossy~\cite{bib:FerrariEtAl-2011a} flooding scheme. In Glossy the flood initiator transmits a frame containing at least a hop counter. When nodes receive that frame, they increment the hop counter and rebroadcast the frame after a fixed small delay. If the time jitter in the frame retransmission is kept below 500ns, 802.15.4 frames interfere constructively~\cite{bib:FerrariEtAl-2011a} and thus it is possible to flood the network without creating a spanning tree and scheduling retransmissions. One important but overlooked characteristic of Glossy is that it provides each node with knowledge of how many hops there are between it and the flood initiator. In TDMH-MAC, where floods are only initiated by the master node, we exploit this information to forward network topology information towards the master node.

The topology collection is performed by having each node periodically broadcast its topology information. For each control uplink slot only one node can broadcast, and the topology information is not flooded, so only its direct neighbors can receive it. Nodes that overhear this frame can update their local knowledge of the network topology. Nodes that cannot reach the master directly -- thus having a hop number higher than 1 -- also randomly select a node with a lower hop number as fowardee of their topology information. This node will store this information and later, when its turn to transmit comes, forward it together with its own data.
This solution exploits the hop information made available by Glossy to guarantee that at every retransmission the topology information is always forwarded closer to the master node, efficiently routing it in the minimum number of transmissions.

The information that each node transmits in its turn is the following: its node ID, its hop, the node ID of the forwardee, and a bitmask with its current knowledge of its direct neighbors.
Bitmasks are fixed size, requiring a number of bytes equal to the maximum number of nodes divided by eight, thus not imposing any limit on the number of neighbors a node can have.
Moreover, a node also transmits forwarded topologies composed of node IDs and bitmasks, as well as requests to open new streams, that will be discussed later on.
The number of forwarded topologies is limited by the control uplink slot size.
In dense networks, a node that is selected as forwardee by more nodes than the available forwarding capability puts forwarded topologies in a queue and sends them in fifo order.
The control uplink message can be set larger than a single IEEE 802.15.4 frame, allowing more forwarding space for large networks.
Finally, if propagation delay compensation is enabled in the network configuration, the same round robin scheme for topology collection is used for propagation delay compensation. In this case, the control uplink frames are considered an implicit propagation delay request by the node, and overhearing nodes of the previous hop reply with the cumulated propagation delay frame~\cite{bib:TerraneoEtAl-2015a}.

To avoid collisions, the node that can transmit in each control uplink slot is selected using a round robin scheme, where the global network time made available from clock synchronization is used to number the slots starting from the maximum number of nodes in the network minus one, counting down towards one, and then repeating. There is no slot reserved for the master node, which only overhears during uplink slots. The choice of a downcounting ordering reduces the time for the algorithm to converge in the common case where nodes with a lower node ID are closer to the master.
Nodes that are no longer overheard for a configurable number of rounds of the algorithm are removed from the topology information, in order to respond to link and node failures.

\begin{figure}[t]
 \begin{center}
  \includegraphics[width=0.65\columnwidth]{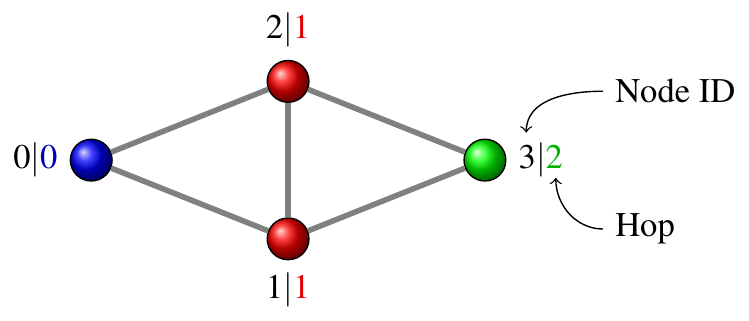}
 \end{center}
 \caption{Network topology used in the topology collection example.}
 \label{fig:ExampleTopology}
\end{figure}

To see how the proposed topology collection algorithm works, consider a newly formed network configured for a maximum of 8 nodes, but having only four. The network topology is shown in Figure~\ref{fig:ExampleTopology}. 
Without loss of generality, consider the situation where all nodes are turned on at the same time.

After the first time synchronization flood, and before the control uplink round robin starts, the knowledge of the network at each node is summarised below (lines indicate known links):

\vspace{1mm}\includegraphics[width=0.95\columnwidth]{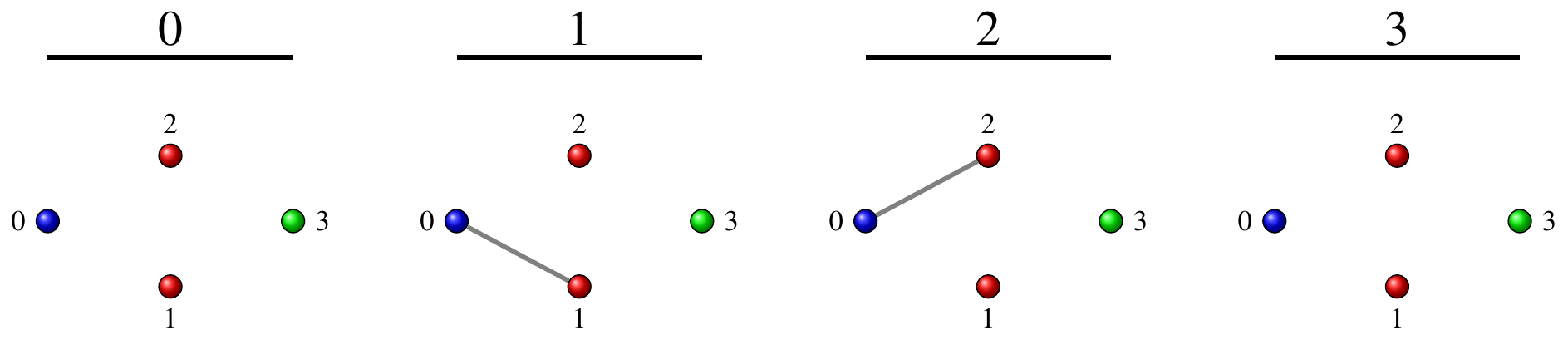}

Nodes 1 and 2 infer that they can reach the master node, since they have gained from the clock synchronization flood that they belong to hop 1. The master node and node 3 have not overheard any topology message, therefore they have no network knowledge.

The first four control uplink slots are reserved for nodes 7, 6, 5 and 4 which do not exist in the network. In this case, all the nodes listen in these time slots, but nothing is received. In the next slot, node 3 transmits the following message: \{nodeID=3, hop=2, forwardee=3, neighbors=$\emptyset$, forwarded=$\emptyset$\} ($\emptyset$ means the empty set). Note that setting $forwrdee = nodeID$ means the node doesn't know any node with a lower hop number, and thus cannot yet forward its topology towards the master.

Nodes 1 and 2 overhear this message, and update their network topology as follows:

\vspace{1mm}\includegraphics[width=0.95\columnwidth]{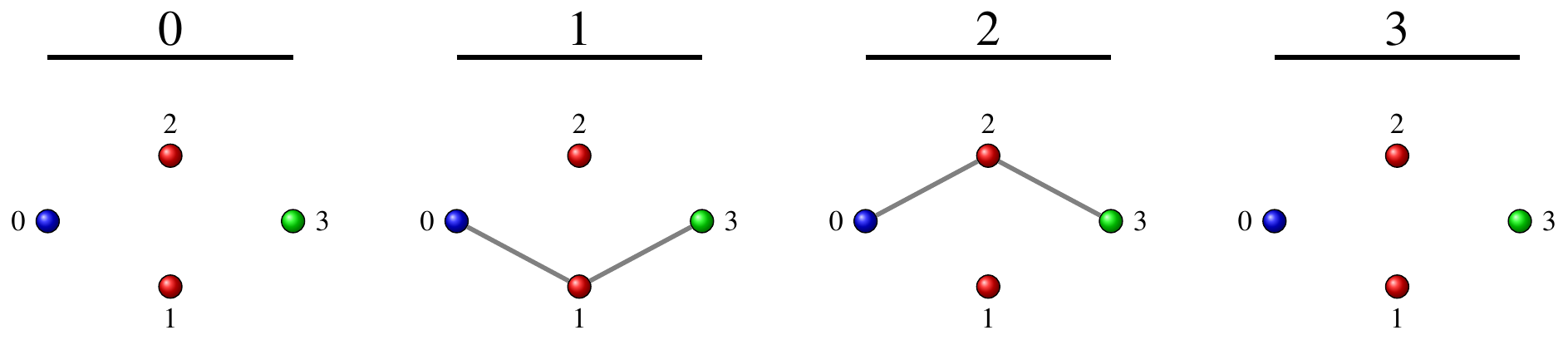}

In the next control uplink slot, node 2 transmits the following message: \{nodeID=2, hop=1, forwardee=0, neighbors=\{0,3\}, forwarded=$\emptyset$\}.

Nodes 0,1 and 3 overhear this message and update their topology information. Node 3 now knows a node with a previous hop number, and the next round will have at least one node to forward its topology to.

\vspace{1mm}\includegraphics[width=0.95\columnwidth]{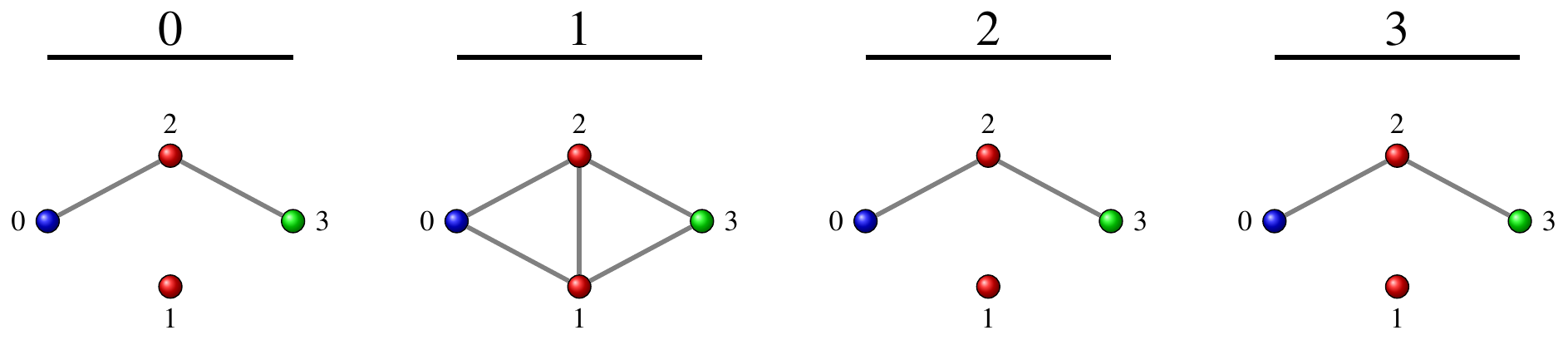}

To conclude the first control uplink round, node 1 transmits its message, which is \{nodeID=1, hop=1, forwardee=0, neighbors=\{0,2,3\}, forwarded=$\emptyset$\}. This message is overheard by nodes 0, 2 and 3 that update their topology as follows:

\vspace{1mm}\includegraphics[width=0.95\columnwidth]{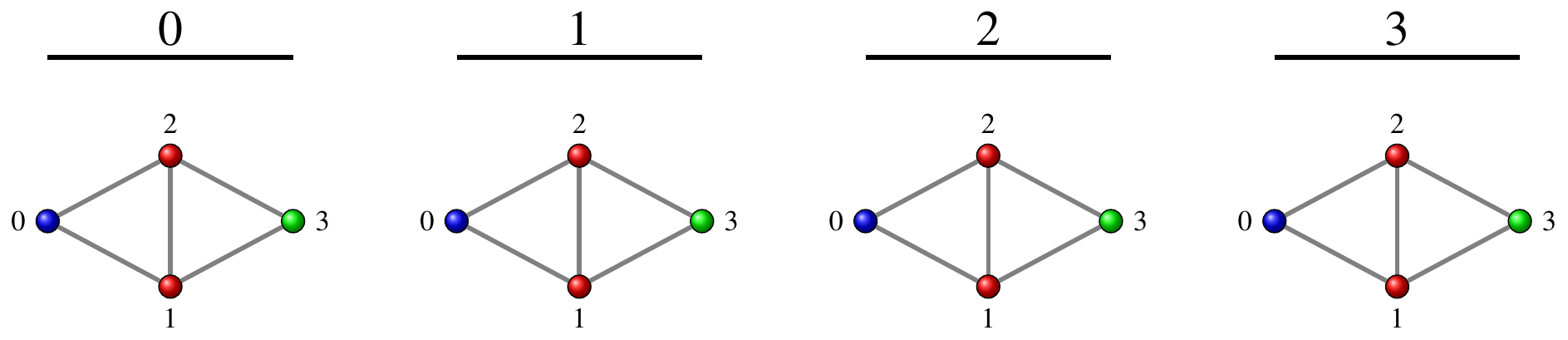}

In this simple example the algorithm converged without the need to forward topologies, and all nodes reached a full knowledge of the network graph, but this is not always the case. For larger networks, only the master node converges to a full network topology. Other nodes may only have a partial network knowledge. This is not a problem for TDMH-MAC, however, since it is centralized.

%% file: sections/05-ScheduleDistribution.tex
\section{Stream scheduling and Schedule Distribution}
\label{sec:Schedule}
TDMH-MAC is mainly targeted at real-time systems performing periodic tasks. Hence, the viewpoint chosen to identify data flows in the protocol deservedly is connection oriented. This choice allows a centralized collision-free  management of the network without excessive control overhead, as stream opening and closing is assumed a more infrequent operation than data transmission, and can be consequently assigned less bandwidth than the data transmission activity.

The communications are managed as logical point to point links called streams, directed from a source to a destination node and unidirectional. A stream supports periodic traffic, composed of a certain number of data packets to be transmitted in a given application-specified period. Streams are opened by the sender node which has to fill a data structure containing the stream information called a \textit{Stream Management Element} (SME) and forward it to the master node.
SMEs are forwarded to the master node like the topology information, in control uplink slots.

The master node is in charge of keeping track of the currently active streams, the current network topology, and has to reschedule the data tansmissions whenever either of the two changes. Contrary to the control superframe, whose length is fixed at the network formation, the length of the data superframe is adapted on-line by the master to accomodate the requested stream periods.
Stream periods can be constrained to be chosen from a given set to make sure the data superframe size does not grow too large to accomodate non harmonic periods.
The computed schedule has to be disseminated through the control downlink flooding. To provide reliability in case of frame corruption, the schedule is sent multiple times before it takes effect. During this time, the previous schedule is executed.

The scheduler and the data format of the schedule distribution frames are not specified in this paper, and are instead considered customization points of the proposed MAC protocol, as also done by other protocols such as TSCH. A simple routing algorithm over the network graph to identify the paths of each stream followed by a greedy heuristic allocation is the minimum required for the MAC to function, but more elaborate solutions could balance the data forwarding load to maximize network lifetime, select multiple paths for streams which request redundancy for additional reliability, or select a schedule minimizing the difference with respect to the previous one in order to use a delta encoding scheme for the schedule dissemination, allowing efficient use of the control uplink bandwidth for large data superframes.

\subsection{Schedule constraints}
\label{sec:scheduler}

Although the scope of this publication doesn't cover the scheduler implementation, the properties of a schedule are here formalized in first order logic, in a view to presenting in a formal way the scheduling problem and establishing a common reference for implementations.

The scheduler will have to satisfy the application requests, which can open and close streams. Each stream consists of periodic transmissions of application-level packets from a source node to a destination node. Since redundancy is provided as a mean to cope with external interference, each application-level packet can map to more than one data frame, each of which travels a path on the topology graph, being forwarded towards the destination node. A stream is therefore composed of a set of individual paths transfering a single data frame along the network graph between the same source and destination.
It is advantageous to formulate the properties of a schedule in terms of the individual data frames and their paths, rather than the streams.

The schedule constraints are first presented in first order logic and then the purpose of each proposition is briefly explained.

\vspace{-0.3em}

\begin{align}
T(i, j, t) :=&\, \text{Transmission from node i to node j at time t}.& \nonumber &\\
G \subset N^2 :=&\, \text{Topology graph}.& \nonumber &\\
P(i, j, p, z) :=&\, \text{Path from node i to node j} & \nonumber\\
&\text{with periodicity p and index z}.& \nonumber &
\end{align}
\vspace{-1.5em}
%
\begin{flalign}
\text{Connectivity} & \label{for:fol:conn}\\
\forall i, j \in N, \forall t, T(i, j, t) \Longrightarrow (i, j) \in G & \nonumber \displaybreak[3] \\
%
\text{Unique sender and receiver} & \label{for:fol:uniq}\\
\forall i,j\in N,\forall t\exists T(i,j,t)\Longrightarrow \forall k,l,u,v\in N|k\neq i,l\neq j & \nonumber \\
\nexists\left(T(u,i,t),T(i,l,t),T(k,j,t),T(j,v,t)\right) & \nonumber \displaybreak[3] \\
%
\text{Contemporary transmissions coexistance} & \label{for:fol:singlerec}\\
\forall i, j, k, l \in N \mid i \neq k, j \neq  l, \forall t \exists T(i, j, t), T(k, l, t) & \Longrightarrow  \nonumber \\
(i, l), (k, j) \notin G & \nonumber \displaybreak[3]\\
%
\text{No transmissions out of a schedule} & \label{for:fol:nospurious}\\
\forall i, j, u, v \in N,\, \forall t, p \in T\, \exists T(i, j, t) & \Longrightarrow  \nonumber \\
\exists P(u, v, p, z) \mid T(i, j, t) \in P(u, v, p, z) & \nonumber \displaybreak[3] \\
%
\text{Path transmission periodicity} & \label{for:fol:periodicity}\\
\forall i, j, u, v \in N,\, \forall t, p \in T \exists T(i, j, t) \in P(u, v, p, z) & \Longleftrightarrow \nonumber \\
\exists T(i, j, t+p) & \nonumber \displaybreak[3] \\
%
\text{Path transmission causality} & \label{for:fol:causality}\\
\forall i, j, k, l \in N,\, \forall t, p \in T\, T(i, j, t) \in P(k, l, p, z) &
\Longleftrightarrow  \nonumber \\
k = i \wedge l = j & \nonumber \\
\vee k = i \wedge \exists u \in N,\, \tau \in \left(t, t+p\right) & \nonumber \\
\wedge T(j, u, \tau) \in P(k, l, p, z) & \nonumber \\
\vee l = j \wedge \exists v \in N,\, \upsilon \in \left(t-p, t\right) & \nonumber \\
\wedge T(v, i, \upsilon) \in P(k, l, p, z) & \nonumber \\
\vee \exists u, v \in N,\, \tau \in \left(t, t+p\right),\, \upsilon \in \left(t-p, t\right) & \nonumber \\
\wedge T(j, u, \tau) \in P(k, l, p, z) \wedge T(v, i, \upsilon) \in P(k, l, p, z) & \nonumber \displaybreak[3] \\
%
\text{Single transmission, single path} & \label{for:fol:singlestream}\\
\forall i, j, u, v, k, l \in N,\, \forall t, p, q \in T,\, T(i, j, t) \in P(u, v, p, z) \wedge & \nonumber \\ 
T(i, j, t) \in P(k, l, q, w) & \Longrightarrow  \nonumber \\
P(u, v, p, z) = P(k, l, q, w) & \nonumber \\
\end{flalign}

A transmission is an event occurring at a given slot number in the data superframe from a source node to a destination node, which must be in the radio range of the transmitting node, hence proposition~\ref{for:fol:conn}.
A node cannot receive transmissions from more than one sender at the same time, cannot transmit to multiple receivers at the same time, and finally a node cannot transmit and receive at the same time (proposition~\ref{for:fol:uniq}).
Since in the current implementation of TDMH-MAC all nodes transmit on the same channel, concurrent transmissions in the same slot are possible only if they do not interfere. For this to occur, there can't exist two sending nodes in the radio range of a receiver node at any given time (proposition~\ref{for:fol:singlerec}).
Proposition~\ref{for:fol:nospurious} forbids spurious transmissions that do not belong to paths.
Proposition~\ref{for:fol:periodicity} simply states that transmissions are periodic.
Proposition~\ref{for:fol:causality} constrains transmissions belonging to a path to be causal, by stating that for each transmission in a path, there must be a preceding one towards the source, and a following one towards the destination, and that they must occur within the prescribed period. Note that the notation $(t, t+p)$ represents an open interval.
Finally, proposition~\ref{for:fol:singlestream} forbids transmissions from belonging to mutiple paths.

%% file: sections/06-DataTransmission.tex
\section{Data Transmission}
\label{sec:DataTransmission}

Each node processes the flooded schedules, which contain the information for all nodes, and extracts the part of the schedule that concerns it. From this information, nodes allocate a certain number of forwarding buffers, that are used to store incoming frames to be later transmitted. To reduce the memory footprint, buffers can be reused for unrelated streams using an algorithm similar to the liveness analysis in programming languages.
Moreover, for streams with redundancy only one buffer can be used, and nodes that successfully receive a frame need not listen during the other data slots for redundancy, to save energy.

Once the local node schedule is computed from the flooded information and buffers are allocated, nodes know in advance what to do in each data slot of a data superframe, and repeatedly perform the same operation in every data superframe until a new schedule takes effect.

This feature enhances energy efficiency, as nodes can sleep for the entire length of unused slots, resulting in an energy consumption that scales linearly with the data rate traversing the node.

%% file: sections/07-SimScalability.tex
\section{Simulation exploration and scalability}
\label{sec:expres}

The reference implementation of TDMH-MAC is available as free software\footnote{https://github.com/fedetft/tdmh}. The MAC has been implemented in C++, a language supported by both the Miosix kernel used in Wandstem~\cite{bib:Terraneo:2016:DHE:2893711.2893753} WSN nodes, as well as the OMNeT++~\cite{omnet} network simulator. By implementing the radio transceiver and time API of the Miosix kernel through a wrapper in OMNeT++, it was possible to have a single codebase that could run both on the simulator and on real WSN nodes.
The implementation in the nodes allows to test the MAC in a real-world scenario, where constructive interference has to occur also in the presence of multipaths, transmissions are affected by interference from other networks such as Wi-Fi, and clock synchronization has to track the time varying drift of quartz crystals.
The availability of a simulator allows to test the protocol also without having access to WandStem nodes, to develop new schedulers, and to test the MAC in large scale deployments.
It should however be noticed that OMNeT++ is not a cycle accurate CPU simulator, and thus the resource constraints both in terms of limited CPU processing power and RAM memory of the real nodes are not enforced by the simulator.

In this section the MAC performance is characterized using simulations and conclusions are drawn in tems of its scalability, while in the next section the actual link and data transmission reliability is tested through experiments.

\subsection{Control overhead}

The necessity to transmit control frames for network management reduces the available bandwidth for data transmission.
TDMH-MAC, thanks to its deterministic operation, allows to compute the control overhead a-priori, with just the network configuration information. This fact allows to guarantee how much bandwidth is available for data transmission, and is one of the features that makes bounded latency communication possible.
This is unfortunately not true in most other MACs, making a comparison difficult.

In TDMH-MAC, all data transmissions follow a TDMA approach where the data slot length is constant.
Thus, with a given data slot length, the efficiency depends on the number of slots used for network control operations over the horizon of a control superframe, which as explained in Section~\ref{sec:Protocol} can be divided in uplink and downlink.

The number of slots required for control downlink is simple, as it depends on the maximum number of hops the network is configured for, as those frames are flooded.

The number of slots required for control uplink instead depends on the number of uplink frames and whether propagation delay compensation is active. The minimum configuration is one uplink frame per control uplink slot.
This is a good configuration for networks with propagation delay compensation disabled and up to around 32 nodes (for larger networks the time for propagation of topology changes would grow too large due to limited space for forwarding topologies). A full configuration would use three frames per control uplink, two of which are used to increase the number of forwarded topologies to handle a larger number of nodes, and the third is the propagation delay compensation reply as specified in~\cite{bib:TerraneoEtAl-2015a}.

\begin{figure}[t]
 \begin{center}
  \includegraphics[width=0.95\columnwidth]{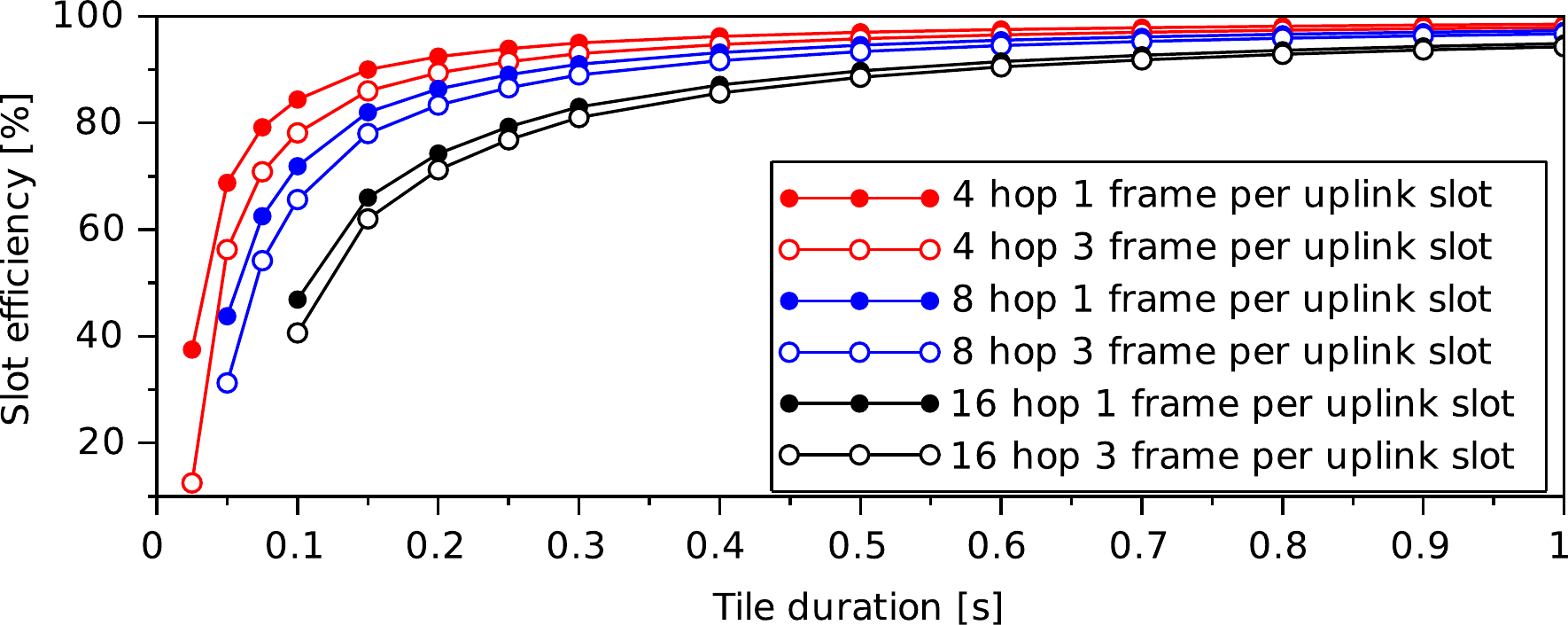}
 \end{center}
 \caption{Percentage of slots usable for data transmission as a function of tile duration and configuration.}
 \label{fig:Efficiency}
\end{figure}

Figure~\ref{fig:Efficiency} shows the percentage of slots available for data transmission as a function of tile duration, with different maximum hop numbers and number of control uplink frames per uplink tile. The figure assumes a configuration where the data frame size is set to 125 Bytes, the maximum supported by the 802.15.4 physical layer
and the data slot has a length of 6 ms.
The figure shows how TDMH-MAC can achieve high slot efficiency with reasonable tile durations, while for networks requiring tracking of fast topology changes a trade-off exists.

\subsection{Power efficiency}

The power consumption of running TDMH-MAC on a WSN node is also very predictable. It can be divided in that caused by data transmission, which can be computed off-line given the current schedule, and that due to control frames exchange, which can be estimated from the network configuration and some topology information.
In detail, the consumption caused by the downlink flooding is simple to estimate, as every node has to receive and rebroadcast exactly one frame per flood, and the flooding period is known in the network configuration. The consumption due to the uplink phase instead depends on the number of neighbors a node has, as nodes that do not sense the start of a frame in the first part of the TDMA slot can go to sleep.

\begin{figure}[t]
 \begin{center}
  \includegraphics[width=0.95\columnwidth]{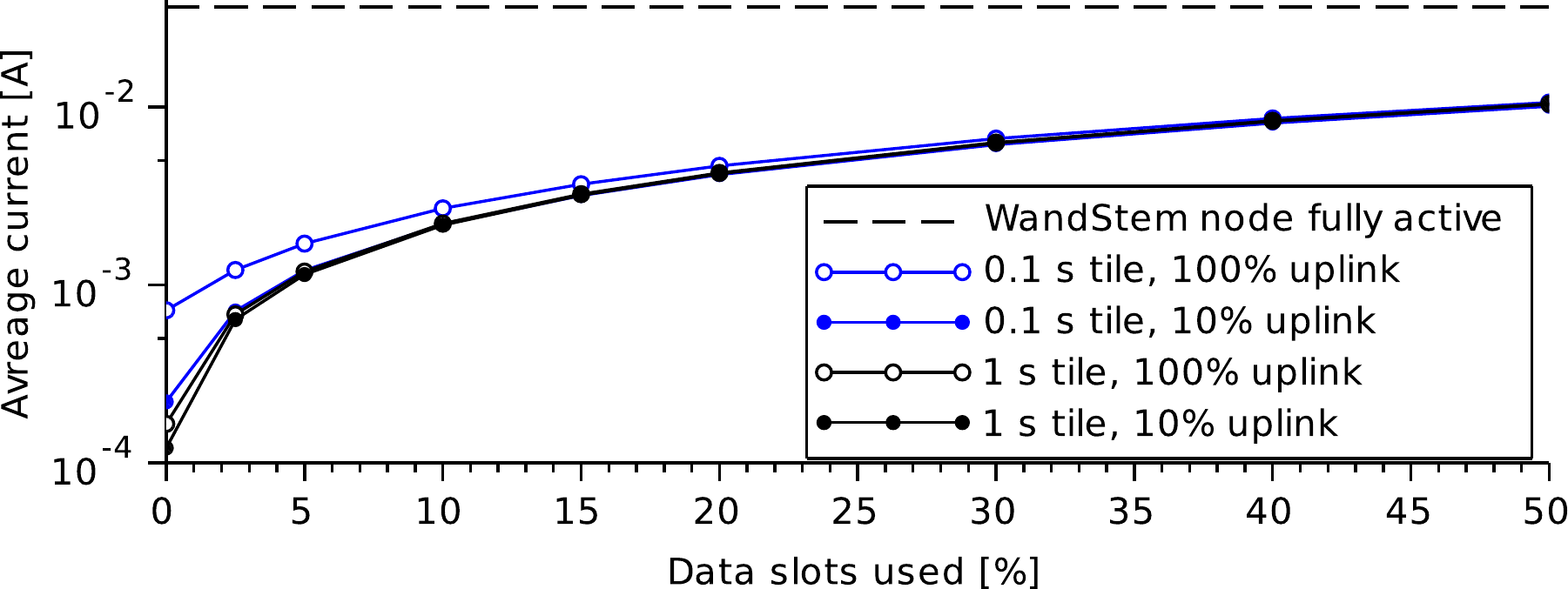}
 \end{center}
 \caption{Average node current consumption as a function of percentage of data slots used by the schedule, tile duration and network connectivity.}
 \label{fig:avgcurrent}
\end{figure}

Figure~\ref{fig:avgcurrent} plots the average current consumption of a node running TDMH-MAC as a function of the percentage of data slots used for data transmission, tile duration and average percentage of uplink slots where a frame is overheard (network connectivity).
The plot clearly shows that the control activity of TDMH-MAC is very efficient, as current consumption is dominated by the data frames transmitted, and the MAC overhead becomes significant only when the data rate is close to zero.
When the data load is exactly zero, for short tile durations and fully connected networks, where in every uplink slot a frame is overheard, the average current consumption can reach 0.72mA. For long tile durations and a sparse network, where in only 10\% of uplink slots a frame is overheard, the consumption can be as low as 0.12mA, most of which is due to the FLOPSYNC-VHT timebase~\cite{bib:TerraneoEtAl-2017a} necessary to keep synchronization while in deep sleep.
When the data slots used grows to just 10\%, the average current in the two aforementioned cases becomes 2.69mA and 2.17mA, showing how the dominant part of the consumption is due to data transmission.

Another advantage of TDMH-MAC is that given its deterministic nature, the root node can estimate the average current consumption of each node from the schedule and network topology.

\subsection{Topology collection convergence}

The convergence time of the topology collection algorithm determines the reaction time of TDMH-MAC to a topology change, such as a node joining or leaving the network, or links becoming available/unavailable due to environmental changes or interference. A special case is the initial network formation problem.

Computing the convergence time to a topology change is difficult to perform in closed form due to the need to take into account the queues of topology messages in every node. The OMNeT++ simulator of TDMH-MAC can however be used to simulate a given network condition and easily compute the convergence time.
The convergence time mainly depends on four factors. The first one is the network configuration. The maximum number of nodes, tile duration and superframe structure determine the round robin period where each node has a chance to broadcast its existence. Moreover, increasing the number of frames per uplink slot can improve convergence time for large networks, where forwarding becomes a bottleneck. The second factor is the number of nodes actually present, whcih is in general different from the maximum number of nodes the network is configured for. Third, the network topology impacts the convergence time, with a star network being the best case, and a line topology the worst. Finally, due to the round robin scheme, the assignment of node IDs in the topology also has an impact, the best case being with nodes with low ID being close to the master, and the worst case being the reverse.

\begin{figure}[tb]
 \begin{center}
  \includegraphics[width=0.45\columnwidth]{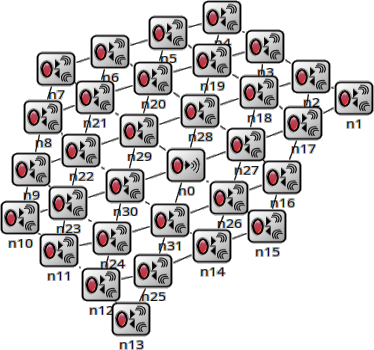}
 \end{center}
 \caption{Example of a 32 node network simulated in OMNeT++.}
 \label{fig:omnet32node}
\end{figure}

Given the high number of factors affecting convergence, and the resulting difficulty to explore the space of possibilities, it was decided to perform two simulation campaigns presenting the relevant case of network formation time and convergence time to a node failure as a function of the network configuration and number of nodes. The other factors were set as follows: the network topology was chosen hexagonal-like, where each node has up to six neighbors. Figure~\ref{fig:omnet32node} shows an example of such a network with 32 nodes. Node IDs were assigned in reverse order, the worst case assignment. Uplink slots are composed of a single frame. Tile duration was set to 100ms, resulting in a control overhead ranging from 7 to 30\%.

\begin{figure}[tb]
 \begin{center}
  \includegraphics[width=0.95\columnwidth]{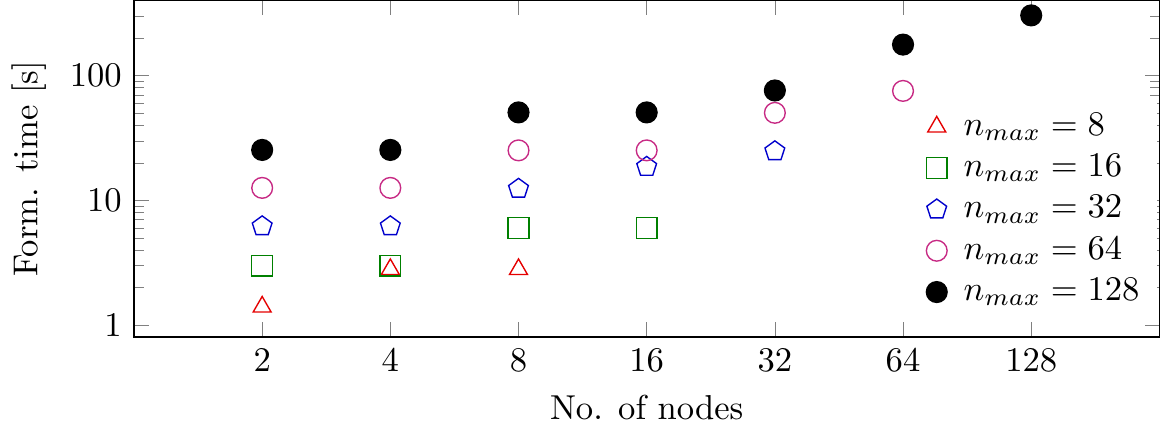}
 \end{center}
 \caption{Network formation time as a function of number of nodes and network configuration.}
 \label{fig:figure_formation_time_sim}
\end{figure}

Figure~\ref{fig:figure_formation_time_sim} shows the network formation time, counted starting when the node clocks are synchronized, and ending when the master has the full graph of the network. From this graph it can be noticed that for networks under 32 nodes the formation time is under 100 seconds. For larger networks it grows up to 304.8s for a network with 128 nodes, limited by topology forwarding. In such a case, increasing the number of frames per uplink slot can be beneficial.

\begin{figure}[tb]
 \begin{center}
  \includegraphics[width=0.95\columnwidth]{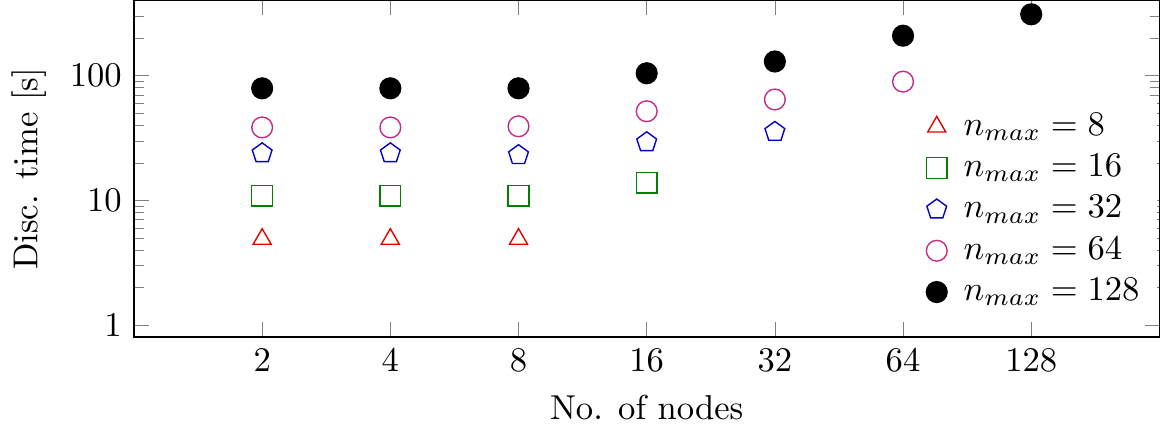}
 \end{center}
 \caption{Network convergence time after node failure as a function of number of nodes and network configuration.}
 \label{fig:figure_disconnect_time_sim}
\end{figure}

Figure~\ref{fig:figure_disconnect_time_sim} shows the convergence time after the failure of the node with ID 1. This node was chosen as it is the farthest away from the master, thus taking into account the need to forward topology changes. As can be seen, the convergence time is comparable to the network formation, but a bit larger because after the node has failed, the topology is updated only if it is not overheard for a configurable number of rounds (three rounds in the simulations). For example, in the case of a network with $n=32$ and $n_{max}=128$, the total disconnect time is 130.1s, but during the first 53.9s the neighbors no longer overhear the node, and then the new topology takes 76.2s to reach the master. 

\subsection{Scalability consideration}

From the simulation presented, it can be concluded that TDMH-MAC can easily scale to networks with more than 100 nodes and 10 hops. Moreover, there are no topology constraints (e.g: on the number of neighbors a node can have).
Thus, it can be safely stated that the efficient topology collection and schedule distribution strategies of TDMH-MAC allowed to extend the state of the art in terms of real-time communication over low data rate networks.
Further scaling would exacerbate the tradeoff between control overhead and topology convergence, as well as increasing the schedule size and consequently introducing lengthy dissemination times. To overcome this tradeoff, a different physical layer than 802.15.4 would be needed, with a higher data rate than 250kbit/s, although supporting constructive interference on high data rate physical layers remains an open research area.
A higher data rate making it possible to reduce the topology convergence time would also allow supporting mobile nodes, even though power consumption may rise significantly.

%% file: sections/08-ExpRes.tex
\section{Experimental evaluation}

\begin{figure}[t]
 \begin{center}
  \includegraphics[width=0.95\columnwidth]{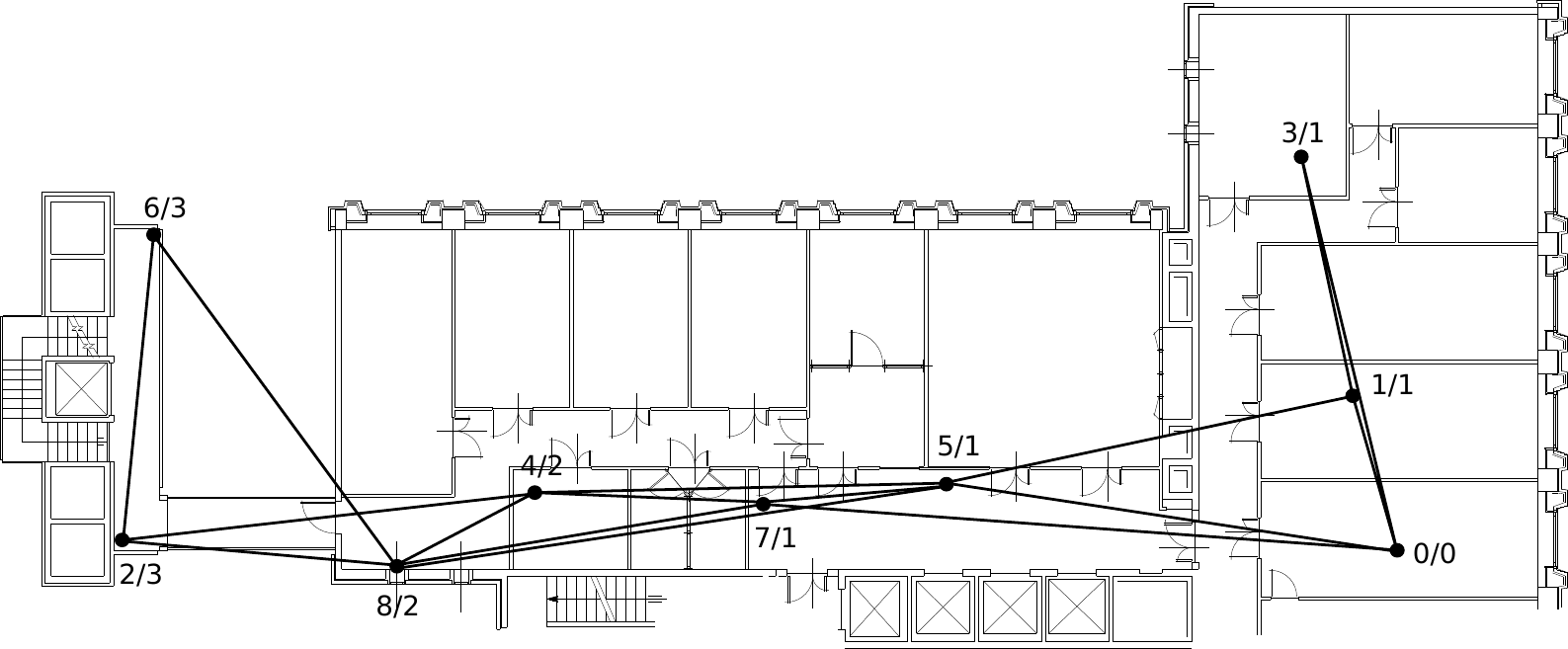}
 \end{center}
 \caption{Node placement for the experimental evaluation, showing the node ID, hop and the network graph produced by the topology collection. Only links with reliability greater than 80\% are shown.}
 \label{fig:ExperimentTopology}
\end{figure}

For the experimental evaluation, nine Wandstem WSN nodes were distributed in the first floor of building 21 of the Politecnico di Milano, as shown in Figure~\ref{fig:ExperimentTopology}. The node location was chosen to represent a generic deployment scenario in an office building and to maximise the number of hops, limited by the available space and the rooms we had access to. Note that the network graph shown is taken from the topology collection algorithm running on the nodes.
For this evaluation the MAC was configured for a maximum of 32 nodes and 6 hops, with a tile duration of 100 ms, and a control superframe with one downlink and one uplink tile. Data frame size was set to 125 Bytes (the maximum allowed by the physical layer) resulting in 6 ms data slots. Since the data slot size does not divide the tile duration, the MAC automatically inserts a 4 ms slack time at the end of each tile where nodes can sleep. Propagation delay compensation was not enabled, since the network spans just 45 m.

We present two experiments. The first one aims at assessing the link reliability in a real-world scenario, and consists of starting the network, letting the topology collection reach a steady state, and then logging the topology during a two days timespan. The used reliability metric is the link uptime (the percentage of time each link is listed in the topology).

\begin{table}[b]
{
    \vspace{0.4em}
    \caption{Link reliabilities for the two-day experiment.}
    \vspace{0.4em}
    \footnotesize
     \begin{center}
      \begin{tabular}{lr|lr|lr}
        Link  & Reliability & Link & Reliability & Link & Reliability \rule[-0.9ex]{0pt}{0pt}\\
        \hline
        0-1   & 100.00\%    & 2-4  &  99.01\%    & 4-7  &  98.83\% \\             
        0-3   & 100.00\%    & 2-6  &  92.58\%    & 4-8  &  97.89\% \\                   
        0-5   & 100.00\%    & 2-7  &   6.03\%    & 5-7  &  99.67\% \\
        0-7   &  99.74\%    & 2-8  &  98.98\%    & 5-8  &  93.52\% \\
        1-3   & 100.00\%    & 3-5  &  55.09\%    & 6-8  &  84.21\% \\
        1-5   &  93.34\%    & 4-5  &  99.27\%    & 7-8  &  96.21\% \\
        1-7   &  56.14\%    & 4-6  &  15.70\%                      \\
      \end{tabular}
      \label{tab:linkreliability}
     \end{center}
}
\end{table}

Results are shown in Table~\ref{tab:linkreliability}. Most reliabilities are above 80\%, while some nodes far apart can sporadically connect, resulting in a few low reliability links. Links with low reliability are due to distance and/or external interference. However they do not pose a threat to TDMH-MAC, thanks to the mesh topology. The scheduler will reschedule streams when links become unavailable, and the spatial redundancy can compensate for link failures until a reschedule occurs.

\begin{figure}[t]
 \begin{center}
  \includegraphics[width=0.95\columnwidth]{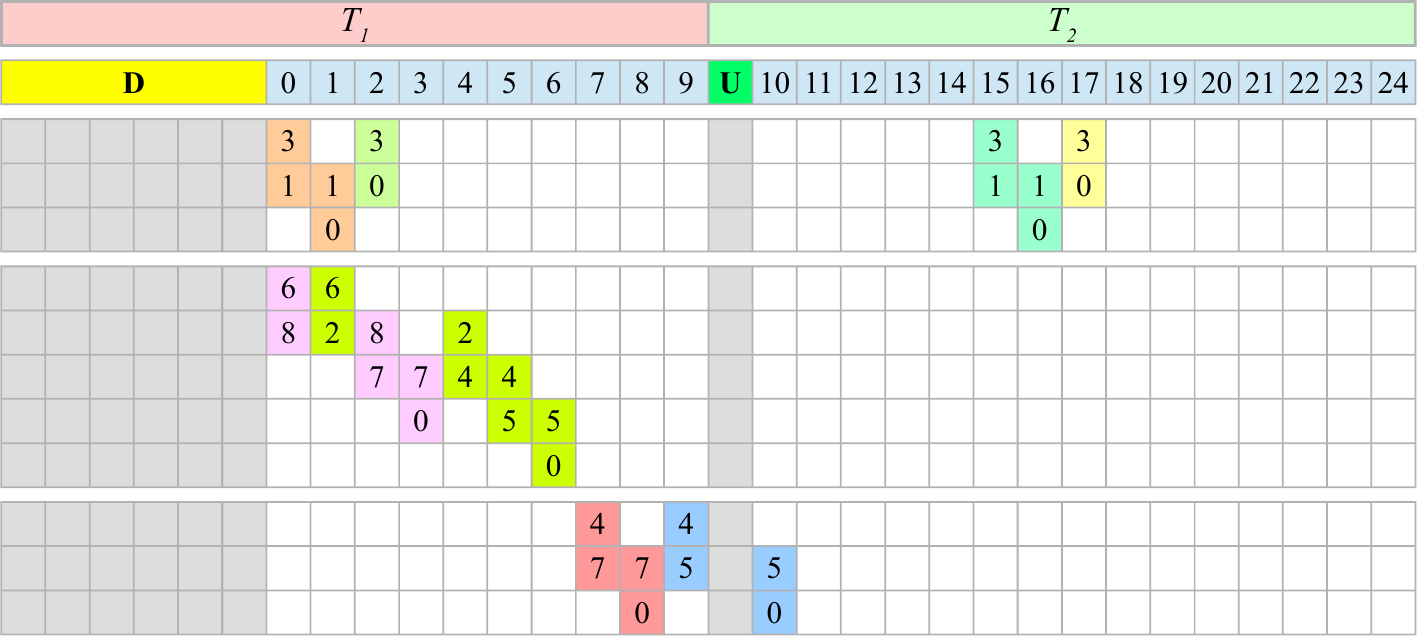}
 \end{center}
 \caption{Schedule with double redundancy. 
X axis shows slots within the 200 ms control superframe (grey columns are control slots).
Y axis groups the three streams 3 $\,\to\,$ 0, 6 $\,\to\,$ 0 and 4 $\,\to\,$ 0. Colored boxes are individual frames being transmitted through their path. Top number is transmitting node, bottom number receiving, color is the path.}
 \label{fig:Schedule}
\end{figure}

The second experiment aims at illustrating redundancy. Three streams were scheduled from node 3, 4 and 6 towards node 0. The stream from 3 to 0 had a 100 ms period, the others had 200 ms. Two tests (each one day long) were made. The first was without redundancy. The second had double spatial redundancy, forcing the duplicated frames across different paths. The used reliability metric is the percentage of application-level packets correctly received. The schedule for the experiment with redundancy is shown in Figure~\ref{fig:Schedule}; note the concurrent transmissions among non-interfering nodes. It can be seen that the stream from node 3 to 0 has a latency bound of 18ms, the one from node 6 to node 0 a latency bound of 42ms, while the stream from node 4 to node 0 has a bound of 30ms. It should however be noted that delay bounds may vary if the topology changes, and a rescehdule is needed. The worst-case delay bound provided by the scheduler is equal to the stream period itself, which is a common assumption in real-time systems.

\begin{table}[hb]
{
    \vspace{0.4em}
    \caption{Stream reliability with and without redundancy.}
    \vspace{0.4em}
    \footnotesize \centering
        \begin{tabular}{lrr}
           
            Stream        & Without redundancy & With redundancy \\
            \hline
            3 $\,\to\,$ 0 & 99.56\%            & 99.67\%         \\
            6 $\,\to\,$ 0 & 95.40\%            & 99.97\%         \\
            4 $\,\to\,$ 0 & 97.45\%            & 99.00\%         \\
        \end{tabular}
        
    \label{tab:datareliability}
}
\end{table}

The results with and without redundancy are shown in Table~\ref{tab:datareliability}: TDMH-MAC can provide reliable data transmission, and this is improved by redundancy.

%% file: sections/09-Conclusions.tex
\section{Conclusions and future work}
\label{sec:conclusions}

In this paper an innovative wireless MAC targeted at real-time applications was proposed. By relying on state-of-the-art low overhead clock syncronization and a previously overlooked hop information made available by constructive interference flooding schemes a distributed algorithm was designed capable of efficiently collecting and forwarding the current network graph to the master node.

This in turn enables a centralized network resource allocation eliminating collisions and enabling bounded latency periodic communication. Constructive interference flooding was also used to efficiently disseminate the network schedule to all nodes in the network.

In TDMH-MAC, network reliability to interferences and link topology changes can be dealt with through not only temporal redundancy by transmitting the same frame multiple times, but also spatially, routing the retransmitted frames through different links and nodes thus taking full advantage of the mesh nature of the network.

An experimental evaluation showed the capabilities of the protocol in a real-world deployment.

It is expected that this protocol will be used as the basis for real-time distributed applications, and the opportunity made available by the scheduler customization point will foster additional research from the real-time community.

The proposed MAC has been implemented in a unitary codebase that can be run both on a widely used network simulator, and on an Open Hardware wireless node platform targeted at real-time low power communication, thereby presenting a low entry barrier for exploitation and additional research.